\documentclass[%
aps,
prl,
reprint,
superscriptaddress,
nofootinbib,
amsmath,
amssymb
]{revtex4-2}
\usepackage{
    physics,
    graphicx,
    multirow,
    tabularx,
    mathtools, 
    placeins, 
    nicefrac, 
    hyperref, 
    threeparttable, 
    siunitx,
    enumitem,
    makecell
}

\usepackage[capitalize]{cleveref}
\Crefname{section}{Sec.}{Secs.}

\usepackage[dvipsnames]{xcolor}
\hypersetup{
    colorlinks,
    linkcolor={blue!50!black},
    citecolor={blue!50!black},
    urlcolor={blue!80!black}
}

\usepackage[caption=false]{subfig} 
\begin{document}


\title{Practical Quantum Advantage for Boosting Citations}

\author{Zhenhuan Liu}
\affiliation{Global Institute for Seeking Quantum Advantage}


\begin{abstract}
Realizing practical quantum advantage with meaningful economic impact is the holy grail of the quantum information field.
Recent quantum technology advances have driven exponential growth in quantum information research, with resultant publications achieving significantly elevated impact.
Within academia, citation counts serve as a key metric for evaluating research impact, often directly influencing career advancement and compensation structures.
Motivated by these observations, we propose a potential protocol for practical quantum advantage in boosting citations.
\end{abstract}

\maketitle
\section{Introduction}
Quantum physics has demonstrated significant advantages over classical physics in various information processing tasks. 
Notable examples include quantum key distribution for secure communication~\cite{bennett1984}, quantum simulation for modeling complex systems~\cite{feynman1982}, and quantum metrology for enhanced precision measurements~\cite{giovannetti2011}. 
In particular, quantum computing has shown the potential for exponential speedups in some classical problems such as integer factorization~\cite{shor1997polynomial}. 
However, current quantum devices still face limitations in coherence time, error rates, and scalability, making it challenging to fully realize these advantages in practice. 
A key focus in quantum information science today is leveraging near-term quantum processors to demonstrate computational advantages in practical problems while also delivering tangible economic value~\cite{arute2019,preskill2018}.

Citations serve as a critical metric for evaluating researchers’ scientific impact, reflecting the adoption and influence of their work within the global academic community.
Highly cited publications not only enhance a researcher’s reputation but also directly contribute to career advancement, funding acquisition, and institutional rankings~\cite{hirsch2005hindex}.
Beyond academia, citations translate into economic value by signaling technological relevance and commercialization potential. 
For instance, foundational papers in fields like genomics, artificial intelligence, or quantum computing often correlate with patents, industry collaborations, and startups~\cite{azoulay2019funding}. 
Studies show that citation counts are strongly linked to research funding return on investment, as they help policymakers and investors identify high-impact areas~\cite{lane2011investment}. 
In emerging fields, citation trends can even guide research and development investment decisions, bridging the gap between theoretical breakthroughs and market-driven applications~\cite{bornmann2014evaluating}. 
Thus, citations function as both an academic currency and an economic indicator, incentivizing innovation with measurable societal returns.

\begin{figure}
\centering
\includegraphics[width=0.85\linewidth]{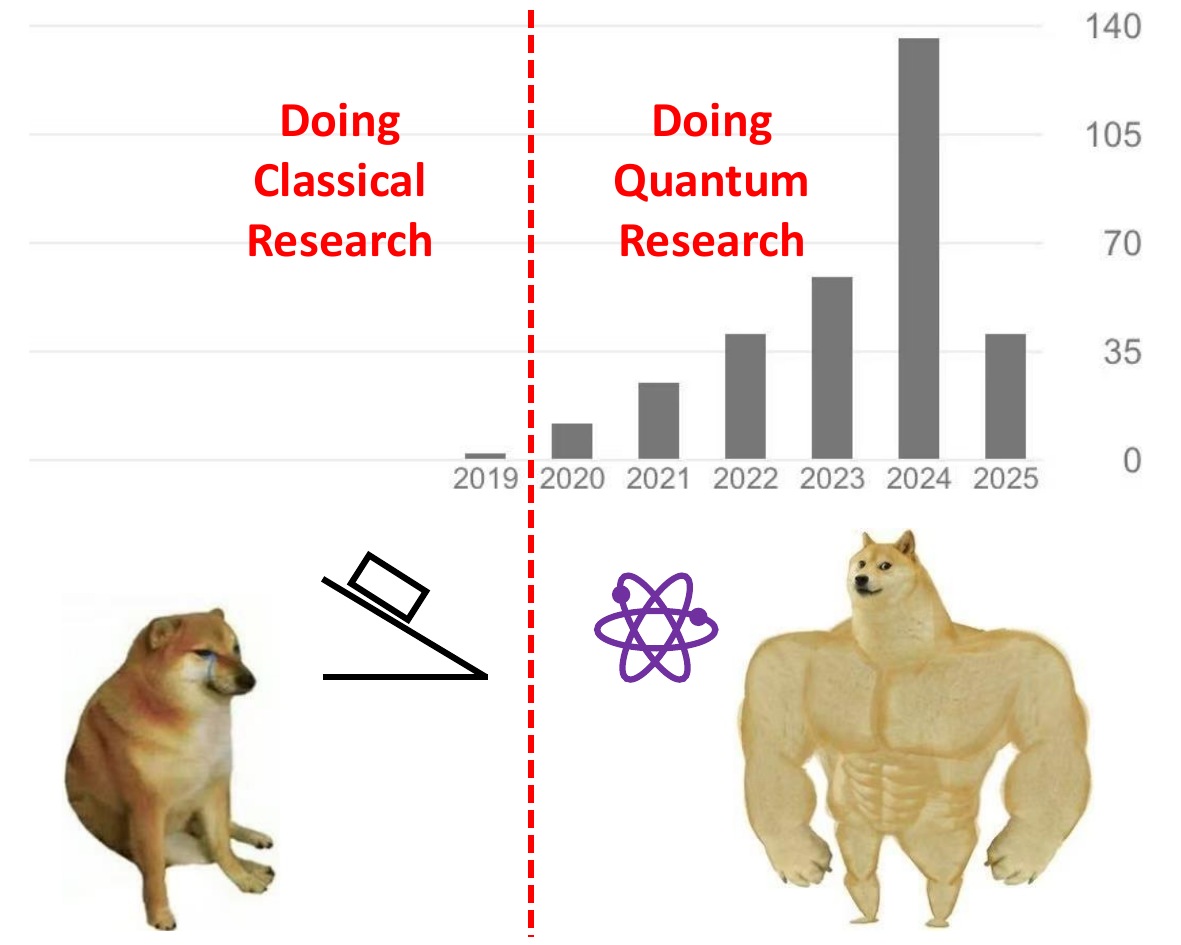}
\caption{Doing quantum research helps to gain exponentially more citations compared with doing classical research! Notice that we are just one-fourth into 2025.}
\label{fig:overview}
\end{figure}

In this work, we propose a novel protocol for demonstrating practical quantum advantage with economic significance. 
As illustrated in Fig.~\ref{fig:overview}, our Quantum Boosting Citations (QBC) protocol requires minimal quantum resources, potentially operating without specialized quantum devices. 
We present a detailed experimental implementation scheme and validate our approach through empirical data analysis. 
These comprehensive investigations establish our proposal as a significant milestone in the pursuit of practical quantum advantages.

\section{Quantum Citation Boosting}
The QCB protocol is solely consisted with three steps:
\begin{enumerate}
\item Do quantum research.
\item Publish papers.
\item Wait for other researchers to cite your papers.
\end{enumerate}
The QCB protocol imposes minimal requirements on quantum devices, as the initial stage of it is not restricted to experimental investigations. 
For theoretical studies, the protocol can be implemented exclusively using classical computing devices (e.g., laptops or brains). 
These characteristics make the QCB protocol particularly suitable for near-term demonstration.

To more effectively demonstrate the exponential advantage of quantum research over classical research in increasing citations, researchers should carefully select their research directions.
First, focusing on trending subfields can be beneficial. These areas typically have more open problems and attract greater research attention, making it easier to gain citations. However, the challenge is that working in popular directions requires continuous learning and staying updated with cutting-edge developments.
Alternatively, researchers may choose to investigate simpler subfields. These areas, being more accessible, often serve as entry points for newcomers to the field. Since the number of early-career researchers far exceeds that of established experts in any field, significant breakthroughs in these fundamental areas can potentially attract substantial citations from this larger audience.

\section{Experiment}
\begin{figure}
\centering
\includegraphics[width=0.9\linewidth]{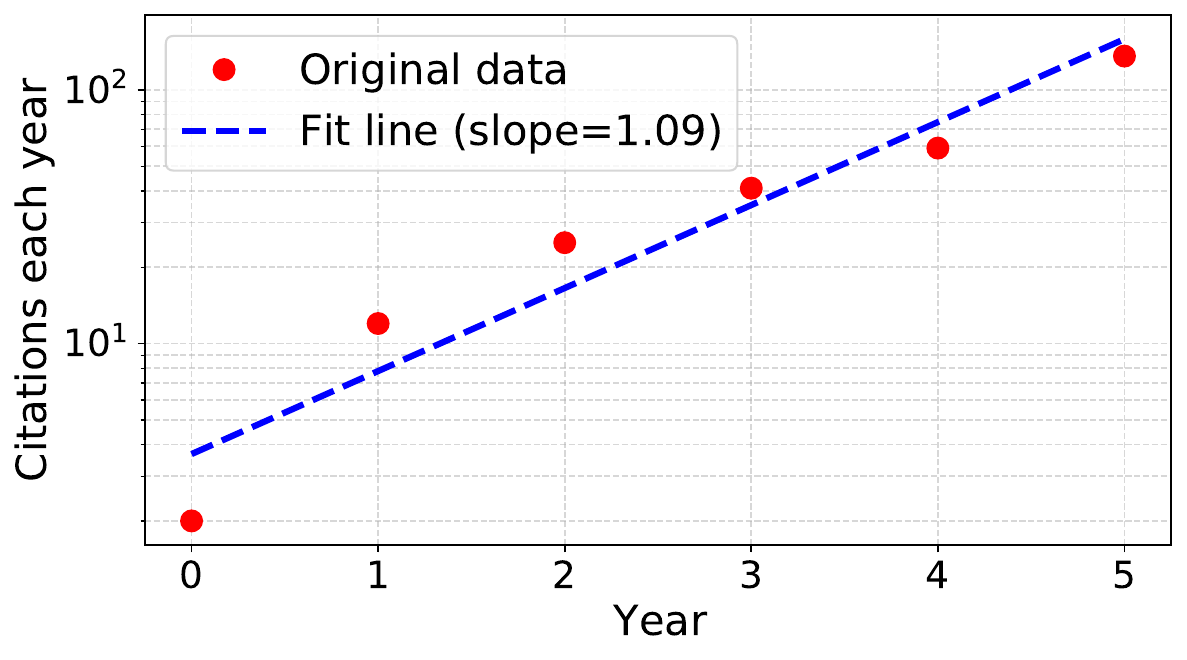}
\caption{The relation between citations with year after 2019.}
\label{fig:fit}
\end{figure}

We randomly (Haar random) select a quantum information researcher to demonstrate the performance of QCB protocol, who was doing classical physics research before 2020 and turned into quantum information research afterwards, as shown in Fig.~\ref{fig:overview}.
It is clear to see that the citation number of this researcher after 2020 has an significant increment, which is exponential with time.
To make our conclusion more convincing, we fit the data of citations per year and time (after 2019) using the exponentiation function, 
\begin{equation}
\mathrm{Citations \ per \ year}=2^{\alpha\times \mathrm{year}+\beta}.
\end{equation}
We show the regression result in Fig.~\ref{fig:fit}, which shows that the citations per year doubles every eleven months.
The coefficient of determination is $R^2=0.982$, showing that the behavior of the data can be effectively described using the exponentiation function.
Therefore, this shows a strong evidence that QBC has an exponential advantages compared with classical strategies.

\section{Discussion and Outlook}
In this work, we propose a novel form of practical quantum advantage for enhancing citation impact. While our experimental results validate this approach, a comprehensive theoretical framework remains to be developed. Several critical questions must be addressed.

First, regarding the universality of the QBC protocol: although we randomly sampled the quantum researcher, potential worst-case scenarios may exist. Specifically, certain researchers might not exhibit the predicted exponential citation growth after doing quantum research. This possibility requires systematic investigation.
Second, we must examine the risk of dequantization—whether classical research strategies could achieve comparable citation enhancement to the QBC protocol. This fundamental question bears on the protocol's quantum essentiality.
Furthermore, the long-term dynamics demand scrutiny. Our current analysis in Fig.~\ref{fig:overview} is limited to a ten-year citation window, prompting the natural question: does QBC's advantage persist over extended timescales?

\section{ACKNOWLEDGMENTS}
We highly appreciate the Deepseek in helping us to writing this paper, which is a very nice Chinese large language model.

%

\end{document}